\documentclass[fleqn,twoside]{article}
\usepackage{espcrc2}

\newcommand{\be}{\begin{equation}}
\newcommand{\ee}{\end{equation}}
\newcommand{\ba}{\begin{eqnarray}}
\newcommand{\ea}{\end{eqnarray}}

\newcommand{\1}{^{-1}}
\newcommand{\dg}{^{\dagger}}
\newcommand{\di}{\mbox{d}\,} 
\newcommand{\e}{\mbox{e}}
\newcommand{\ga}{\gamma_5}
\newcommand{\h}{\frac{1}{2}}
\newcommand{\Id}{\mbox{1\hspace{-.9mm}l}}

\newcommand{\mo}[1]{^{(\mbox{\scriptsize #1})}}
\newcommand{\na}[1]{\nabla_{#1}}
\newcommand{\ra}{\rightarrow}
\newcommand{\re}[1]{(\ref{#1})}
\newcommand{\vp}{\varphi} 
\newcommand{\W}{_{\mbox{\scriptsize W}}} 
 
\newcommand{\Tr}{\mbox{Tr}}

\newcommand{\sy}{\scriptscriptstyle} 
\newcommand{\T}{{\cal T}} 
\newcommand{\G}{{\cal G}} 

\newcommand{\AmS}{{\protect\the\textfont2
  A\kern-.1667em\lower.5ex\hbox{M}\kern-.125emS}}

\title{\vspace*{-6mm}
\raisebox{0.8cm}[0pt][0pt]{\makebox[0pt][l]{\parbox{16cm}{\normalsize%
\mbox{}\hfill HUB-EP-02/33}}}\\
More chiral lattice fermions\thanks{
Contribution to Lattice 2002, Cambridge, MA, USA.}}

\author{Werner Kerler \address{Institut f\"ur Physik, 
        Humboldt-Universit\"at, D-10115 Berlin, Germany}%
}
       
\begin{document}

\begin{abstract}
Instead of the Ginsparg-Wilson relation only generalized chiral symmetry is
required. The resulting much larger class of Dirac operators for massless
fermions is investigated and a general construction for them is given. It is 
also shown that the new class still leads properly to Weyl fermions and to
chiral gauge theories. 
\vspace{1pc}
\end{abstract}

\maketitle
\thispagestyle{empty}

\section{INTRODUCTION}

The condition for generalized chiral symmetry of L\"uscher \cite{lu98i}
in Ref.~\cite{lu98} has the general form
\be
\ga D + D \ga V = 0\;, \quad V\dg=V^{-1}=\ga V \ga\;.
\label{gg}
\ee
We only impose \re{gg} together with $\ga$-hermiticity
\be
D\dg=\ga D\ga
\label{g5}
\ee
on $D$ \cite{ke02}. With \re{g5} we get from \re{gg}
\be
D+D\dg V=0 \;,\quad D\dg+DV\dg=0\;,
\label{dd}
\ee
which implies $[V,D]=0$ and $DD\dg=D\dg D$. This is accounted for by
requiring $D$ to be of form 
\be
D=F(V)\;.
\ee

The simplest special case is that of the operators satisfying the 
Ginsparg-Wilson (GW) relation $\;\{\ga,D\}=\rho^{-1}D\ga D$ \cite{gi82}, for 
which one has
\be
D=F(V)=\rho(1-V)\;.
\ee
A further special case is that of the operators of Fujikawa \cite{fu00} 
satisfying $\{\ga, D\}=2D(\ga a_0D)^{2k+1}$ with $k=0,1,2,\ldots$,
for which we find the form 
\be
D=F(V)= a_0\1\,\Big(\h(1-V)(-V)^k\Big)^{1/(2k+1)}.
\ee

\section{SPECTRAL REPRESENTATION}

The spectral representation of $D=F(V)$ in terms of that of $V$ on the finite
lattice is 
\begin{displaymath}
D=f(1)(P_1^{(+)}+P_1^{(-)})+f(-1)(P_2^{(+)}+P_2^{(-)})
\end{displaymath}
\be
\;+\sum_{k\;(0<\vp_k<\pi)}\Big(f(\e^{i\vp_k})P_k\mo{I}+f(\e^{-i\vp_k})
P_k\mo{II}\Big)\;,
\label{SPEC}
\ee
where the orthogonal projections satisfy 
\be
\ga P_j^{(\pm)}=P_j^{(\pm)}\ga= \pm P_j^{(\pm)},\, \ga P_k\mo{I}=
P_k\mo{II}\ga\;. 
\ee
For the functions $f(v)$ from \re{gg} and \re{g5} one gets 
\be
f(v)+f(v)^*v=0\;,\quad f(v)^*=f(v^*)\;,
\label{cond0}
\ee
so that in particular $f(1)=0$ and $f(-1)$ real. In addition to determining 
the form of $D$, the functions $f(v)$ describe the location of its spectrum.

According to the projection properties we have
\begin{displaymath}
\mbox{Tr}(\ga P_1^{(\pm)})=\pm N_{\pm}(1)\;,\quad
\mbox{Tr}(\ga P_2^{(\pm)})=
\end{displaymath}
\be
\pm N_{\pm}(-1)\;,\quad
\mbox{Tr}(\ga P_k\mo{I})=\mbox{Tr}(\ga P_k\mo{II})=0\;,
\label{PR}
\ee
where $N_{+}(\pm1)$ and $N_{-}(\pm1)$ are the dimensions of the right-handed 
and the left-handed eigenspaces of $V$ for eigenvalues $\pm1$, respectively.
Evaluating the expressions, 
\be 
\lim_{\zeta\rightarrow 0}\mbox{Tr}\big(\ga\frac{-\zeta}{D-\zeta\Id} \big)\;,
\quad\lim_{\zeta\rightarrow 0}\mbox{Tr}\big(\ga\frac{D}{D-\zeta\Id} \big)\;,
\label{EX}
\ee
which sum to zero, it follows that in any case
\be  
N_+(1)-N_-(1)+N_+(-1)-N_-(-1)=0
\label{SUM}
\ee
and, in addition, that to allow for a nonvanishing index of $D$, which is 
given by the first expression in \re{EX}, one must require
\be
f(-1)\ne0\;.
\label{cond1}
\ee
With this in place \re{SUM} corresponds to the sum rule found by Chiu 
\cite{ch98} in the GW case. 

Inserting the spectral representation of $V$ into $\mbox{Tr}(\ga V)$ and
using \re{PR} and \re{SUM} we generally obtain for the index of $D$
\be
N_+(1)-N_-(1)=\h\mbox{Tr}(\ga V)\;,
\label{ANO}
\ee
which obviously involves only $V$ for the whole new class of operators $D\,$.
Previous results in the GW case \cite{lu98,ha98} and in the overlap formalism
\cite{na93} can be recognized as special cases of \re{ANO}.

\section{CONSTRUCTION OF $D$}

It follows from \re{cond0} that $f$ has the form 
\begin{displaymath}
f(\e^{i\vp})=\e^{i(\vp-\pi)/2}\,g(\vp)\;,\quad g(\vp)\;\mbox{real}\;,
\end{displaymath}
\be
g(\vp)=-g(-\vp)\;,\quad g(\vp+2\pi)=-g(\vp)\;.
\ee
Thus $D=F(V)$ can be obtained by determining functions $g(\vp)$ with
the above properties. 

The immediate choice for such $g$ is the set
\be
g(\vp)=\sum_{\nu}s_{\nu}w_{\nu}(t_1,t_2,\ldots) 
\ee 
with real functions $w_{\nu}$ and
\be
s_{\nu}=\sin(2\nu+1)\frac{\vp}{2}\;,\quad t_{\mu}=\cos\mu\vp\;.
\ee
Because of the identity $s_{\nu}=s_0 (1+2\sum_{\mu=1}^{\nu}t_{\mu})$
and since the $t_{\nu}$ can be expressed by polynomials of $t_1$
this can be reduced to the simpler form 
\be
g(\vp)=s_0\,w(t_1)=\sin\!\frac{\vp}{2}\;w(\cos\vp)\;.
\label{g0}
\ee
Given a function $g$ with the required properties then $h(g)$ again has these
properties provided that $h$ satisfies for real $x$ 
\be
h(-x)=-h(x)\;,\quad h(x)^*=h(x)\;.
\ee
Thus instead of \re{g0} we get the more general form
\be
g(\vp)=a\1h\big(\sin\!\frac{\vp}{2}\;w(\cos\vp)\big)\;,
\ee
including the factor $a\1$ for later convenience.

To satisfy condition \re{cond1} we need $g(\pi)\ne0$ or $h\big(w(-1)\big)\ne0$.
Therefore we have to impose
\be
w(-1)\ne0\;,
\ee
which is sufficient if $h(x)$ gets only zero for $x=0\,$. To guarantee this 
we require strict monotony,
\be
h(y)>h(x)\;\;{\rm for}\;\;y>x\;.
\ee
Then also the inverse function $\eta$ with 
\be
h\big(\eta(x)\big)=x
\ee
is uniquely defined, which we need below.

Inserting the obtained construction of $f$ into the spectral representation 
\re{SPEC} we have
\be
D=\frac{1}{ia}V^{\h}H\Big(\frac{1}{2i}(V^{\h}-V^{-\h})\;W\big(\textstyle{\h}
(V+V\dg)\big)\Big)
\label{DG}
\ee
where the properties of the hermitian operator functions $W$ and $H$ are 
determined by those of the real functions $w$ and $h$, respectively. 

To specify $V$ appropriately we generalize the overlap form of $V$ \cite{ne98} 
introducing
\be
V=-D\W^{(\eta)}\Big(\sqrt{D\W^{(\eta)\dag}D\W^{(\eta)}}\,\Big)\1\;,
\ee
\begin{displaymath}
D\W^{(\eta)}=iE\Big(\frac{1}{2i}\sum_{\mu}\gamma_{\mu}(\na{\mu}-\na{\mu}
\dg)\Big) 
\end{displaymath}\vspace*{-5mm}
\be
\qquad\qquad+\;E\Big(\frac{r}{2}\sum_{\mu}\na{\mu}\dg\na{\mu}\Big)+
E\big(m\Id\big)\;, 
\ee \vspace*{-2mm}
where the properties of the hermitian operator function $E$ (of hermitian 
operators) are determined by those of the real function $\eta$ (of real $x$). 

To check the continuum limit we note that for the Fourier transform of $V$ in 
the free case one has at the corners of the Brillouin zone $\tilde{V}=-1$ and 
at zero
\be 
\tilde{V}\ra 1-\frac{i}{|\eta(m)|}\,\tilde{E}\Big(a
\sum_{\mu}\gamma_{\mu}p_{\mu}\Big)\;\;{\rm for}\;\;a\ra0\;.
\ee
Requiring $\tilde{W}(-1)\ne0\,$, because of the monotony of $E(X)$ doublers are
suppressed for $-2r<m<0$ as usual. Since $H(E(X))=X$, putting
$\tilde{W}(1)=2|\eta(m)|$ the correct limit of the propagator is obtained 
generally for the class of $D$ of form \re{DG}. 

\section{CHIRAL GAUGE THEORIES}

The projection operators given in \cite{lu98} and implicit in \cite{na93} are
of the general form 
\be
P_{\pm}=\h(1\pm\ga)\Id\;,\quad \tilde{P}_{\pm}=\h(1\pm \ga V)\Id\;,
\label{PP}
\ee
in which only $V$ is involved. The latter also holds for the difference of 
dimensions
\be
\Tr\,P_+-\Tr\,\tilde{P}_- =\h\Tr(\ga V)\;,
\ee
which equals \re{ANO}. 

According to \re{gg} we have $D=\h(D-\ga D\ga V)$, 
which with \re{PP} becomes $D=P_+D\tilde{P}_-+P_-D\tilde{P}_+\,$. With this 
we generally get for Weyl operators 
\be
P_{\pm}D\tilde{P}_{\mp}= D\tilde{P}_{\mp}= P_{\pm}D\;.
\ee 

The projections can be represented in terms of bases with 
$u_i\dg u_j=\delta_{ij}$, $\tilde{u}_k\dg\tilde{u}_l=\delta_{kl}$ as
\be
P_+=\sum_ju_ju_j\dg\;,\quad\tilde{P}_-=\sum_k\tilde{u}_k\tilde{u}_k\dg\;,
\ee \vspace*{-1mm}
which implies that one has the eigenequations
\be
\quad P_+u_j=u_j\;,\quad\tilde{P}_-\tilde{u}_k=\tilde{u}_k\;.
\label{EIG}
\ee 

Associating Grassmann variables $\bar{\chi_j}$ and $\chi_k$ to
the degrees of freedom gives the fields 
\be
\bar{\psi}=\sum_j\bar{\chi_j}u_j\dg\;,\quad \psi=\sum_k\tilde{u}_k\chi_k\;,
\ee
and for $\Tr\,\tilde{P}_-=\Tr\,P_+$ for correlation functions
\begin{displaymath}
\int\prod_l(\di\bar{\chi_l}\di\chi_l)\;\e^{-\bar{\psi}D\psi}
\psi_{n({\sy 1})}\bar{\psi}_{n(r_1)}\ldots
\psi_{n({\sy f})}\bar{\psi}_{n(r_f)}
\end{displaymath}\vspace*{-5mm}
\begin{displaymath}
\quad=\sum_{s_1,\ldots,s_f}\epsilon_{s_1s_2\ldots s_f}\;
(\tilde{P}_-D\1 P_+)_{n({s_1})n(r_1)}\ldots 
\end{displaymath}
\be
\qquad\qquad\ldots(\tilde{P}_-D\1 P_+)_{n({s_f})n(r_f)}\;\det M\;,
\ee
where the matrix $M$ and its inverse are given by
\be
M_{jk}=u_j\dg D\tilde{u_k}\;,\quad (M\1)_{kl}=\tilde{u_k}\dg D\1 u_l\;.
\ee 

Obviously for the derivations here  generalized chiral symmetry of $D$ has been 
sufficient and the GW relation has {\it not} been needed. This will also hold
in the following Section.

\section{INVARIANCE OF DETERMINANT}

We note that the transformations of interest in fermion space are closely 
related. For unitary $\T$ with $D'=\T D\T\dg$ because of \re{dd} we also have 
$V'=\T V\T\dg$ and vice versa. Then, requiring $\T$ to satisfy
also  $[\ga,\T]=0\,$, with $V'=\T V\T\dg$ because of \re{PP} we also get 
$\tilde{P}_-'=\T\tilde{P}_-\T\dg\,$ and vice versa. Therefore putting 
$\T=\exp\G$, where 
\be
\G\dg=-\G\;, \quad [\ga,\G]=0\;,
\ee
we obtain for the related variations 
\be
\delta V= [\G,V],\;\; \delta D= [\G,D],\;\;
\delta \tilde{P}_-=[\G,\tilde{P}_-].
\label{DEL}
\ee

Varying the second relation in \re{EIG} gives
\be
(\delta\tilde{P}_-)\tilde{u}_k=(\Id-\tilde{P}_-)\delta\tilde{u}_k\;.
\ee
and inserting the last relation of \re{DEL} into this
\be
\tilde{P}_+\delta\tilde{u}_k=\tilde{P}_+\G\tilde{u}_k\;.
\ee
Thus unfortunately only a relation for $\tilde{P}_+\delta\tilde{u}_k$ follows
here while one for $\tilde{P}_-\delta\tilde{u}_k$ would be needed  to
evaluate the last term of \re{CC0} below.

For the gauge variation $\delta\ln\det M =\Tr (M\1\delta M)$ of the chiral
determinant one gets
\begin{displaymath}
\sum_{k,l} (M\1)_{kl}\delta M_{lk}=
\end{displaymath}\vspace*{-6mm}
\be
\qquad\qquad\Tr(\tilde{P}_-D\1\delta D)+\sum_k\tilde{u_k}\dg\delta\tilde{u_k}
\label{CC0}
\;,
\ee\vspace*{-1mm}
which requiring gauge invariance has to vanish.
The first contribution, giving the gauge anomaly, turns out to involve 
generally only $V$,
\be
\Tr (\tilde{P}_-D\1\delta D)=\h\Tr(\G\ga V)\;.
\label{CC1}
\ee

\end{document}